\documentclass[11pt,a4paper]{article}

\usepackage{jheppub}
\usepackage{bbold}

\makeatletter
\def\@fpheader{Prepared for submission to JCAP}
\makeatother

\newcommand{\mn}{{\mu\nu}}

\newcommand{\Mt}{M_{\rm tot}}

\newcommand{\be}{\begin{equation}}
\newcommand{\ee}{\end{equation}}

\makeatletter
\DeclareRobustCommand{\rcite}[1]{%
  \rcite@aux#1,\@nil{#1}%
}
\def\rcite@aux#1,#2\@nil#3{%
  \if\relax#2\relax
    Ref.~\cite{#3}%
  \else
    Refs.~\cite{#3}%
  \fi
}
\makeatother

\title{On stars, galaxies and black holes in massive bigravity}

\author[a,b]{Jonas Enander}
\author[a,b]{Edvard M{\"o}rtsell}
\affiliation[a]{Oskar Klein Centre, Stockholm University,\\Albanova University Center\\ 106 91 Stockholm, Sweden}
\affiliation[b]{Department of Physics, Stockholm University\\
AlbaNova University Center\\ 106 91 Stockholm, Sweden}
\emailAdd{enander@fysik.su.se}
\emailAdd{edvard@fysik.su.se}
\abstract{In this paper we study the phenomenology of stars and galaxies in massive bigravity. We give parameter conditions for the existence of viable star solutions when the radius of the star is much smaller than the Compton wavelength of the graviton. If these parameter conditions are not met, we constrain the ratio between the coupling constants of the two metrics, in order to give viable conditions for e.g. neutron stars. For galaxies, we put constraints on both the Compton wavelength of the graviton and the conformal factor and coupling constants of the two metrics. The relationship between black holes and stars, and whether the former can be formed from the latter, is discussed. We argue that the different asymptotic structure of stars and black holes makes it unlikely that black holes form from the gravitational collapse of stars in massive bigravity.}


\begin{document}

\maketitle
 
\section{Introduction}

The Hassan-Rosen theory (also called massive bigravity or bimetric theory) -- which is the ghost free, non-linear theory of two interacting spin-2 fields -- has garnished a lot of attention concerning its phenomenological applications since it was introduced in Refs.~\cite{2012JHEP...02..126H,Hassan:2011ea} in 2012. In particular, studies of cosmological expansion histories \cite{vonStrauss:2011mq, Volkov:2011an, Comelli:2011zm, Akrami2013JHEP...03..099A, Akrami:2013pna, Nersisyan:2015oha}, structure formation \cite{Comelli:2012db,Khosravi:2012rk,Berg:2012kn,Fasiello:2013woa,Konnig:2013gxa,Konnig:2014dna,
Comelli:2014bqa,DeFelice:2014nja,Solomon:2014dua,Konnig:2014xva,Lagos:2014lca,Enander:2015vja,
Konnig:2015lfa,Aoki:2015xqa,Mortsell:2015exa}, tensor modes \cite{Cusin:2014psa,Johnson:2015tfa,Amendola:2015tua,Fasiello:2015csa}, integrated Sachs-Wolfe effect \cite{Enander:2015vja} and galactic lensing \cite{Enander:2013kza} have been performed. The possibility to couple both fields that are present in the theory to matter has been explored in  Refs.~\cite{Akrami:2013ffa,Huang:2015yga,2015GReGr..47.1838A,deRham:2014naa,deRham:2014fha,Enander:2014xga,Schmidt-May:2014xla,Comelli:2015pua,Gumrukcuoglu:2015nua,Heisenberg:2014rka,Gumrukcuoglu:2014xba,Heisenberg:2015iqa,Lagos:2015sya}. In this paper, we continue the phenomenological investigations of the Hassan-Rosen theory through a study of static and spherically symmetric (SSS) spacetimes describing, e.g., galaxies, stars and black holes.

Since the Hassan-Rosen theory contains two metrics, the SSS spacetimes necessarily become more involved. In this paper we are interested in metrics that are asymptotically flat. These spacetimes are asymptotically classified according to the relative strength of the massive and massless spin-2 mode that the theory contains \cite{Hassan:2012wr}, and the conformal relationship between the two metrics at infinity. Concerning the black hole solutions, it was shown in Ref.~\cite{Deffayet:2011rh} that if one assumes non-singular solutions, the two metrics must share a common Killing horizon. This means that black hole solutions are highly restricted. For star solutions, one has the option of how to couple matter to the two metrics. In this paper we opt for the commonly chosen approach of coupling only one of the metrics to matter. The theory predicts that including a gravitational source gives rise to a fixed relationship between the asymptotic massive and massless spin-2 modes \cite{Babichev:2013pfa,Enander:2013kza}. In this paper we show that this relationship is not the same as that for black holes. This makes it unlikely that the black holes that the theory contains are end-states of the gravitational collapse of matter. A possible cause is that the symmetry between the two metrics that the black holes display through the common Killing horizon, is broken when one couples only one of the metrics to matter.

Spherically symmetric systems in the context of the Hassan-Rosen theory were first studied in Ref.~\cite{Comelli:2011wq}, where, in particular, the perturbative solutions to the equations of motion was published. Ref.~\cite{Volkov:2012wp} performed an extensive numerical study, and gave conditions for the existence of asymptotically flat black hole solutions. These solutions were further studied in depth in Ref.~\cite{Brito:2013xaa}. Star solutions and the so-called Vainshtein mechanism, described further below, was studied in Ref.~\cite{Babichev:2013pfa}. This reference is central to the analysis performed in this paper. Solutions for charged black holes were found in Ref.~\cite{Babichev:2014fka}, and for rotating black holes in Ref.~\cite{Babichev:2014tfa}. Stability properties of the black holes were investigated in  Refs.~\cite{Babichev:2013una,Brito:2013wya,Brito:2013yxa,Babichev:2014oua,Kobayashi:2015yda}. A general review of black holes in massive bigravity can be found in Ref.~\cite{Babichev:2015xha}. 

The goal of this paper is to investigate what conditions on the parameters of the theory that give rise to phenomenologically viable SSS solutions. Allowing for general parameter values by approaching the regime where the Hassan-Rosen theory becomes equivalent to general relativity, we constrain the ratio between the coupling constants of the two metrics. Furthermore, we analyse the relationship between black hole and star solutions, in order to see whether the gravitational collapse of stars can lead to the black hole solutions of massive bigravity.

This paper is organized as follows. In Section~\ref{sec:setup} we introduce the Hassan-Rosen theory and the spacetime configuration under consideration. Section~\ref{sec:asymp} describes the asymptotic solutions, and in Section~\ref{sec:stars} we state the solution for stars and their phenomenology. In Section~\ref{sec:galaxies} we constrain the phenomenology of galaxies. Section~\ref{sec:vacuum} discusses the relationship between stars and black holes, and if black holes in massive bigravity can be considered as end-states of the gravitational collapse of stars. We conclude in Section~\ref{sec:conc}.

We use units where $G=c=1$ and $M_g^2=(8\pi)^{-1}$.

\section{Setup}
\label{sec:setup}

The Lagrangian for the Hassan-Rosen theory is given by
\begin{align}
\label{eq:HRaction}
\mathcal{L}=&-\frac{M_{g}^{2}}{2}\sqrt{-\det g}R_{g}-\frac{M_{f}^{2}}{2}\sqrt{-\det f}R_{f}\nonumber \\
&+m^{4}\sqrt{-\det g}\sum_{n=0}^{4}\beta_{n}e_{n}\left(\sqrt{g^{-1}f}\right)+\sqrt{-\det g}\mathcal{L}_\mathrm{m},
\end{align}
where $\mathcal{L}_m$ is the matter Lagrangian and $e_n$ are the elementary symmetric polynomials presented e.g. in Ref.~\cite{Hassan:2011vm}. Varying the Lagrangian yields the equations of motion
\begin{align}
G^g_{\mu\nu}+m^{2}\sum_{n=0}^{3}\left(-1\right)^{n}\beta_{n}g_{\mu\lambda}Y_{\left(n\right)\nu}^{\lambda}\left(\sqrt{g^{-1}f}\right)&=\frac{1}{M_g^2}T_{\mu\nu},\\
G^f_{\mu\nu}+\frac{m^2}{\kappa}\sum_{n=0}^{3}\left(-1\right)^{n}\beta_{4-n}f_{\mu\lambda}Y_{\left(n\right)\nu}^{\lambda}\left(\sqrt{f^{-1}g}\right)&=0.
\end{align}
Here, we have defined 
\be
\kappa\equiv \left(\frac{M_f}{M_g}\right)^2,
\ee
and the matrices $Y_n$ are given in Ref.~\cite{Hassan:2011vm}. The parameter $\kappa$ is in principle redundant, since it can be put to unity through a rescaling of $f_\mn$ and the $\beta_n$ (see e.g. Refs.~\cite{Hassan:2011vm, Akrami:2015qga}). We will keep it explicit, however, since it makes the limit to general relativity manifest.

For the fields $g_\mn$ and $f_\mn$, we use the following spherically symmetric and diagonal ansatz\footnote{For a non-diagonal ansatz, the equations of motion constrain the solution to be identical to the Schwarzschild-AdS/dS metric, as shown in Ref.~\cite{Comelli:2011wq,Volkov:2012wp}.}
\begin{align}\label{eq:ansatz}
ds_{g}^{2}=&-Q^2dt^{2}+N^{-2} dr^2+r^2d\Omega^2, \nonumber \\
ds_{f}^{2}=&-a^2dt^2+\frac{U^{\prime 2}}{Y^2}dr^2+U^2d\Omega^2,
\end{align}
where a prime signifies a derivative with respect to $r$. This form for $g_\mn$ and $f_\mn$ is the most general diagonal form of the metrics after using the possibility of doing a rescaling of the radial coordinate. Notice that $f_\mn$ can equivalently be written
\be
ds_{f}^{2}=-a^2dt^2+Y^{-2}dU^2+U^2d\Omega^2,
\ee
and $U(r)$ be interpreted as the radial coordinate for the $f$-metric.

The energy density and pressure are given by $\rho(r)=-T^0_0$ and $P(r)=T^i_i/3$ (summation over $i$ implied), and they satisfy the following conservation equation:
\be\label{eq:consP}
P'=-\frac{Q'}{Q}\left(P+\rho\right).
\ee

In this paper, we will combine analytic and numerical studies. For the numerical analysis, we follow Ref.~\cite{Volkov:2012wp} and put the equations of motion in the following form:
\be
\label{eq:numsetup}
\begin{cases}
N'=\mathcal{F}_{1}\left(r,Q,N,Y,U,\rho,P,c,m^2,\beta_{1},\beta_{2},\beta_{3},\kappa\right),
\\
Y'=\mathcal{F}_{2}\left(r,Q,N,Y,U,\rho,P,c,m^2,\beta_{1},\beta_{2},\beta_{3},\kappa\right),
\\
U'=\mathcal{F}_{3}\left(r,Q,N,Y,U,\rho,P,c,m^2,\beta_{1},\beta_{2},\beta_{3},\kappa\right),
\\
Q'=\mathcal{F}_{4}\left(r,Q,N,Y,U,\rho,P,c,m^2,\beta_{1},\beta_{2},\beta_{3},\kappa\right),
\\
P'=\mathcal{F}_{5}\left(r,Q,N,Y,U,\rho,P,c,m^2,\beta_{1},\beta_{2},\beta_{3},\kappa\right),
\end{cases}
\ee
where $c$ is defined below. The function $a$ can be solved for directly once the other fields are given. When $\rho=P=0$, i.e. in vacuum, $\mathcal{F}_1$, $\mathcal{F}_2$, $\mathcal{F}_3$ become independent of $Q$. In vacuum, one thus first solves three first order equations for $N$, $Y$ and $U$, and then integrate $\mathcal{F}_4$ to get $Q$. When $\rho$ and $P$ are non-vanishing, the five first order differential equations instead have to be solved simultaneously.

\section{Asymptotic structure}
\label{sec:asymp}

Since we are interested in solutions that are asymptotically flat, the metrics should approach
\be\label{eq:flatsol}
g_{\mu\nu}\rightarrow \eta_{\mu\nu},\qquad f_{\mu\nu}\rightarrow c^{2}\eta_{\mu\nu},
\ee
at infinity. Here $c$ is an asymptotic conformal factor between the two metrics. In order for Eq.~\ref{eq:flatsol} to be an solution, we need to impose
\begin{align}
\label{eq:b0b4}
\beta_{0}=&-3\beta_{1}c-3\beta_{2}c^{2}-\beta_{3}c^{3},\\
\beta_{4}=&-\beta_{1}c^{-3}-3\beta_{2}c^{-2}-3\beta_{3}c^{-1},
\end{align}
to cancel the cosmological constant terms for $g_\mn$ and $f_\mn$.\footnote{Notice that in Refs.~\cite{Volkov:2012wp} and \cite{Brito:2013xaa} the parametrization 
\be
\beta_{n}=\left(-1\right)^{n+1}\left(\frac{1}{2}\left(3-n\right)\left(4-n\right)-\left(4-n\right)\alpha_{3}-\alpha_{4}\right)\nonumber
\ee
was used, for which $c=1$.}

Linearizing around the flat space backgrounds, i.e. expanding the metric components as  $Q=1+\delta Q$, $N=1+\delta N$, $a=c(1+\delta a)$, $U=cr\left(1+\delta U\right)$ and $Y=1+\delta Y$, gives
\begin{align}\label{eq:linsolQ}
\delta Q =&-\frac{C_{1}}{2r}+\frac{C_{2}\kappa c^{2}}{r}e^{-m_{g}r},\\
\delta N =&-\frac{C_{1}}{2r}+\frac{C_{2}\kappa c^2\left(1+m_{g}r\right)}{2r}e^{-m_{g}r},\\
\delta a =& -\frac{C_{1}}{2r}-\frac{C_{2}}{r}e^{-m_{g}r},\\
\delta Y =& -\frac{C_{1}}{2r}-\frac{C_{2}\left(1+m_{g}r\right)}{2r}e^{-m_{g}r},\\
\delta U =&\frac{\left(1+\kappa c^{2}\right)C_{2}\left(1+m_{g}r+m_{g}^{2}r^{2}\right)}{2m_{g}^{2}r^{3}}e^{-m_{g}r}.\label{eq:linsolU}
\end{align}
These solutions, first appearing in Ref.~\cite{Comelli:2011wq}, are well-known and have been presented on several occasions in the literature. The parameters $C_1$ and $C_2$ regulate the strength of the massive and massless modes. The graviton mass $m_{g}$ is given by
\begin{equation}
m_{g}^{2}=m^{2}\left(1+\frac{1}{\kappa c^{2}}\right)\left(\beta_{1}c+2\beta_{2}c^2+\beta_{3}c^{3}\right).
\end{equation}

Let us discuss the free parameters that we have at our disposal. Of the five $\beta_n$, two have been fixed in order to yield asymptotically flat solutions. This leaves $\beta_1$, $\beta_2$ and $\beta_3$ as free theory parameters. $m^2$ is not a free parameter since it can be absorbed into the $\beta_n$:s. We will keep it explicit, however, since it sets an overall length scale when the $\beta_n$:s are of order unity. As mentioned above, $\kappa$ is also redundant since it can be put to unity through a rescaling of $f_\mn$ and the $\beta_n$:s. Since it is important for discerning solutions that lie close to those of general relativity, we will, however, keep it explicit. Added to this, we have the conformal factor $c$. On the whole, for vacuum solutions, we have four global parameters, the three $\beta_i$:s and $c$, together with the local parameters $C_1$ and $C_2$, which controls the strength of the massless and massive modes. As discussed later, including a gravitational source fixes the relation between $C_1$ and $C_2$.

The equation of motion have the property that under the rescaling 
\begin{align}
&N(r)\rightarrow N(\lambda r), \quad Y(r)\rightarrow Y(\lambda r),\quad  Q(r)\rightarrow Q(\lambda r),\quad a(r)\rightarrow a(\lambda r),\nonumber \\
&U(r)\rightarrow \frac{1}{\lambda} U(\lambda r),\quad \rho\rightarrow \frac{\rho}{\lambda^2},\quad P\rightarrow \frac{P}{\lambda^2},\quad  m\rightarrow \frac{m}{\lambda},
\end{align}
a solution is mapped onto a new solution \cite{Volkov:2012wp}. We will interchangeable use $r_V$ (defined below) or $\lambda_g\equiv m_g^{-1}$ as radial coordinate.

The linear solutions are valid up to the radius where higher order terms become important. This radius is usually called the Vainshtein radius, and was first identified in Ref.~\cite{1972PhLB...39..393V} in 1972. In massive bigravity, the Vainshtein radius is
\be
r_V\equiv \left(\frac{2\Mt}{m_g^2}\right)^{1/3},
\ee
where $\Mt$ is defined as the total mass of a source. In Ref.~\cite{1972PhLB...39..393V}, Vainshtein also conjectured that there should exist a mechanism, later dubbed the Vainshtein mechanism, that effectively restores general relativity inside the Vainshtein radius. That this exists in the context of massive bigravity for SSS spacetimes was shown in Ref.~\cite{Volkov:2012wp} for the case of $\kappa\rightarrow 0$, and in Ref.~\cite{Babichev:2013pfa} for the $r\ll \lambda_g$ limit. It is important to note, however, that the existence of the Vainshtein mechanism depends on the specific choice of the $\beta_i$ parameters. 

For recent phenomenology concerning the Vainshtein mechanism, see Refs.~\cite{Koyama:2015oma,Avilez-Lopez:2015dja,Mortsell:2015exa} and references therein.

\section{Stars}
\label{sec:stars}

In this section we study the phenomenology of stars in massive bigravity. As a source, we use a star with constant energy density $\rho_\star$, pressure $P(r)$ and radius $r_\star$. The pressure has to satisfy the conservation equation~(\ref{eq:consP}), and vanish at the surface of the star. The mass interior to $r$ is 
\be\label{eq:Mr}
M(r)\equiv \frac{1}{2M_g^2}\int_0^r\rho(\tilde{r})\tilde{r}^2 d\tilde{r},
\ee
and the total mass of the star is thus
\be
M_\star = \frac{\rho_\star r_\star^3}{6M_g^2}.
\ee
We have three effective scales for the stars: $r_\star$, $r_V$ and $\lambda_g$. We will assume that $r_\star\ll \lambda_g$, and comment on both the $r_\star <r_V$ scenario as well as $r_\star > r_V$. 

As shown in Refs.~\cite{Enander:2013kza,Babichev:2013pfa}, the introduction of a source fixes the relation of $C_1$ and $C_2$ in the linear solutions to 
\begin{equation}
C_2=-\frac{2C_1}{3},
\end{equation}
and $C_1=2\Mt/(1+\kappa c^2)$.
The linear solutions then become
\begin{eqnarray}
\delta Q&=&-\frac{\Mt}{r(1+\kappa c^2)}\left[1+\frac{4\kappa c^2}{3}e^{-m_g r}\right],\\
\delta N&=&-\frac{\Mt}{r(1+\kappa c^2)}\left[1+\frac{2\kappa c^2}{3}e^{-m_g r}\right],\\
\delta a&=&-\frac{\Mt}{r(1+\kappa c^2)}\left[1-\frac{4}{3}e^{-m_g r}\right],\\
\delta Y&=&-\frac{\Mt}{r(1+\kappa c^2)}\left[1-\frac{2}{3}(1+m_g r)e^{-m_g r}\right],\\
\delta U&=&-\frac{2\Mt[1+m_g r+(m_g r)^2]e^{-m_g r}}{3r (m_g r)^2}.
\end{eqnarray}
Asymptotically, the fields thus look like a massless general relativity (GR) like term plus a Yukawa term. They exhibit the usual vDVZ-discontinuity \cite{vanDam:1970vg,Zakharov:1970cc,Iwasaki:1971uz} which can be probed observationally. As $r\gg \lambda_g$, the Yukawa term decays, however, and the fields look identical to general relativity. Is is only when $r\lesssim \lambda_g$, or when higher order terms become important, that we can expect any observational signatures.

When massive bigravity is used for cosmological applications, for $\kappa c^2\sim 1$, we expect $\lambda_g$ to be of the order of the Hubble scale today. It is then an excellent approximation that $r_\star \ll \lambda_g$. For this framework, it was shown in Ref.~\cite{Babichev:2013pfa} that it is possible to obtain approximative analytical solutions by assuming that all fields and their derivatives are close to the flat space background, with the exception of $U/r$. Defining\footnote{Note that the definitions of $\alpha$ and $\beta$ are generalized compared to Ref.~\cite{Babichev:2013pfa} in which $\beta_1c+2\beta_2c^2+\beta_3c^3$ was normalized to unity.}
\begin{eqnarray}
\mu&\equiv& \frac{U}{cr}-1,\nonumber \\
\alpha&\equiv&-\frac{\beta_2c^2+\beta_3c^3}{\beta_1c+2\beta_2c^2+\beta_3c^3},\nonumber \\
\beta&\equiv&\frac{\beta_3c^3}{\beta_1c+2\beta_2c^2+\beta_3c^3},
\end{eqnarray}
the metric perturbations can be expressed as
\begin{eqnarray}
r\delta Q^\prime &=&\frac{P\left(r\right)r^{2}}{2M_g^2}+\frac{M\left(r\right)}{r}-\frac{\kappa c^2 m_{g}^{2}r^{2}}{2\left(1+\kappa c^{2}\right)}\left(\mu-\frac{\beta}{3}\mu^{3}\right), \label{eq:deltaQp} \\
\delta N &=&-\frac{M\left(r\right)}{r}-\frac{\kappa c^{2} m_{g}^{2}r^{2}}{2\left(1+\kappa c^{2}\right)}\left(\mu-\alpha\mu^{2}+\frac{\beta}{3}\mu^{3}\right)
 , \label{eq:deltaN} \\
r\delta a^\prime &=&\frac{m_{g}^{2}r^{2}\left(r+r\mu\right)'}{2\left(1+\kappa c^{2}\right)\left(1+\mu\right)^{2}}\left[\mu+2\mu^{2}+\frac{1}{3}\left(2-2\alpha-\beta\right)\mu^{3}\right],\\
\delta Y &=&\frac{m_{g}^{2}r^{2}}{2\left(1+\mu\right)\left(1+\kappa c^{2}\right)}\left[\mu+\left(1-\alpha\right)\mu^{2}+\frac{1}{3}\left(1-\alpha+\beta\right)\mu^{3}\right] .
\end{eqnarray}
These fields are thus functions of $M(r)$, $P(r)$ and $\mu$, where
$\mu$ satisfies a seventh-degree polynomial:
\begin{align}\label{eq:poly}
3\left(1+\kappa c^{2}\right)\mu+6\left(1+\kappa c^{2}\right)\left(1-\alpha\right)\mu^{2}+
 \nonumber \\
\frac{1}{3}\left[6\left(1+\kappa c^{2}\right)\alpha^{2}-2\left(17+18\kappa c^{2}\right)\alpha+4\left(1+\kappa c^{2}\right)\beta+10+9\kappa c^{2}\right]\mu^{3}+
 \nonumber \\
\frac{2}{3}\left[6\left(1+\kappa c^{2}\right)\alpha^{2}-\left(7+9\kappa c^{2}\right)\alpha+4\left(1+\kappa c^{2}\right)\beta+1\right]\mu^{4}+
 \nonumber \\
\frac{1}{3}\left[2\left(1+3\kappa c^{2}\right)\alpha^{2}-\left(1+\kappa c^{2}\right)\beta^{2}+2\left(1+2\kappa c^{2}\right)\beta-4\alpha\beta-2\alpha\right]\mu^{5}
 \nonumber \\
-\frac{2}{3}\kappa c^{2}\beta^{2}\mu^{6}-\frac{1}{3}\kappa c^{2}\beta^{2}\mu^{7}\nonumber \\
=-\frac{\left(1+\kappa c^{2}\right)\left(1+\mu\right)^{2}}{m_g^{2}}\left[\frac{2M\left(r\right)}{r^3}\left(1-\beta\mu^{2}\right)-\frac{P\left(r\right)}{M_g^2}\left(1-2\alpha\mu+\beta\mu^2\right)\right].
\end{align}
The function $\mu$ satisfies $-1<\mu\le 0$ for all physically relevant cases.

\begin{figure}
\begin{centering}
\includegraphics[scale=0.5]{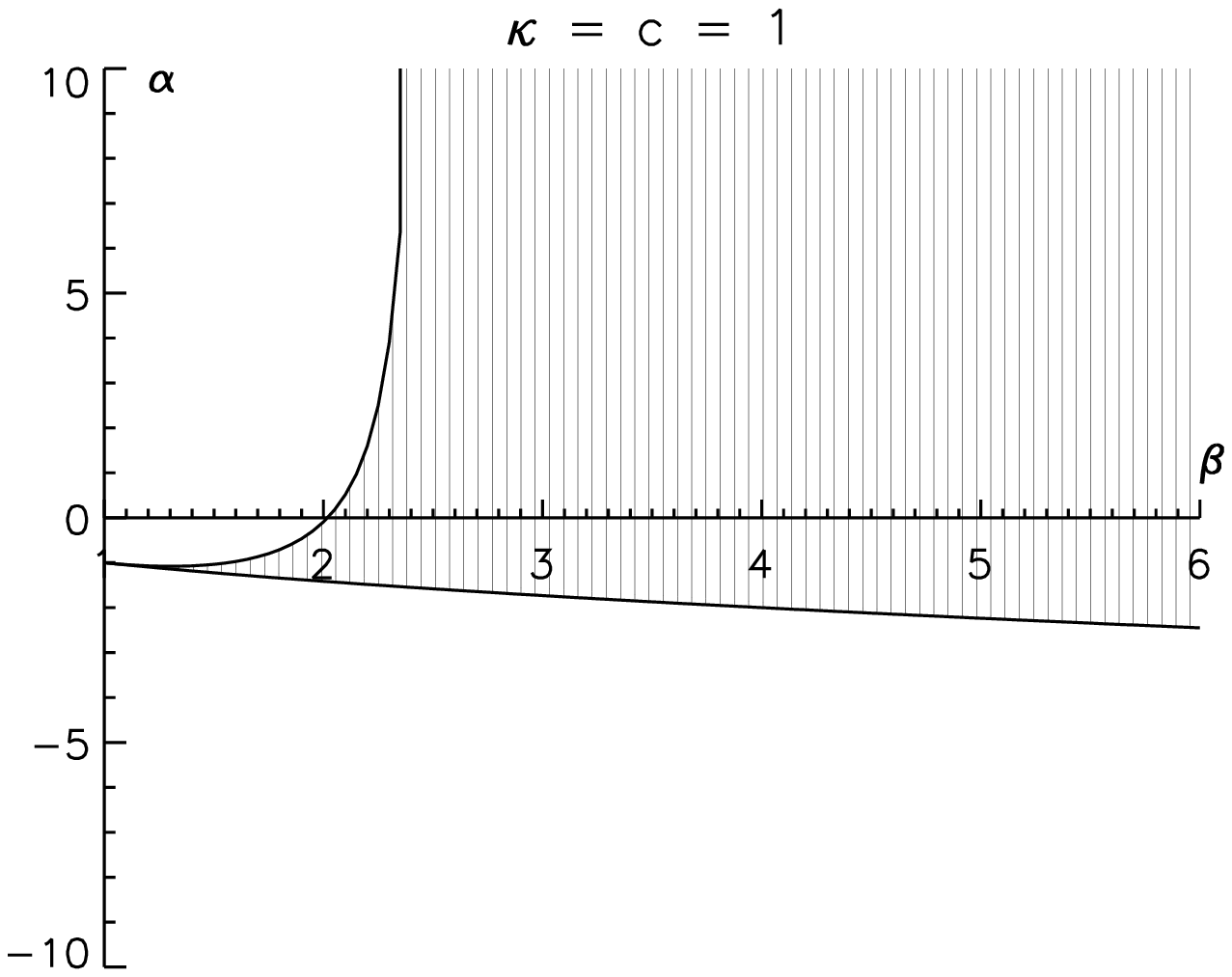}
\includegraphics[scale=0.5]{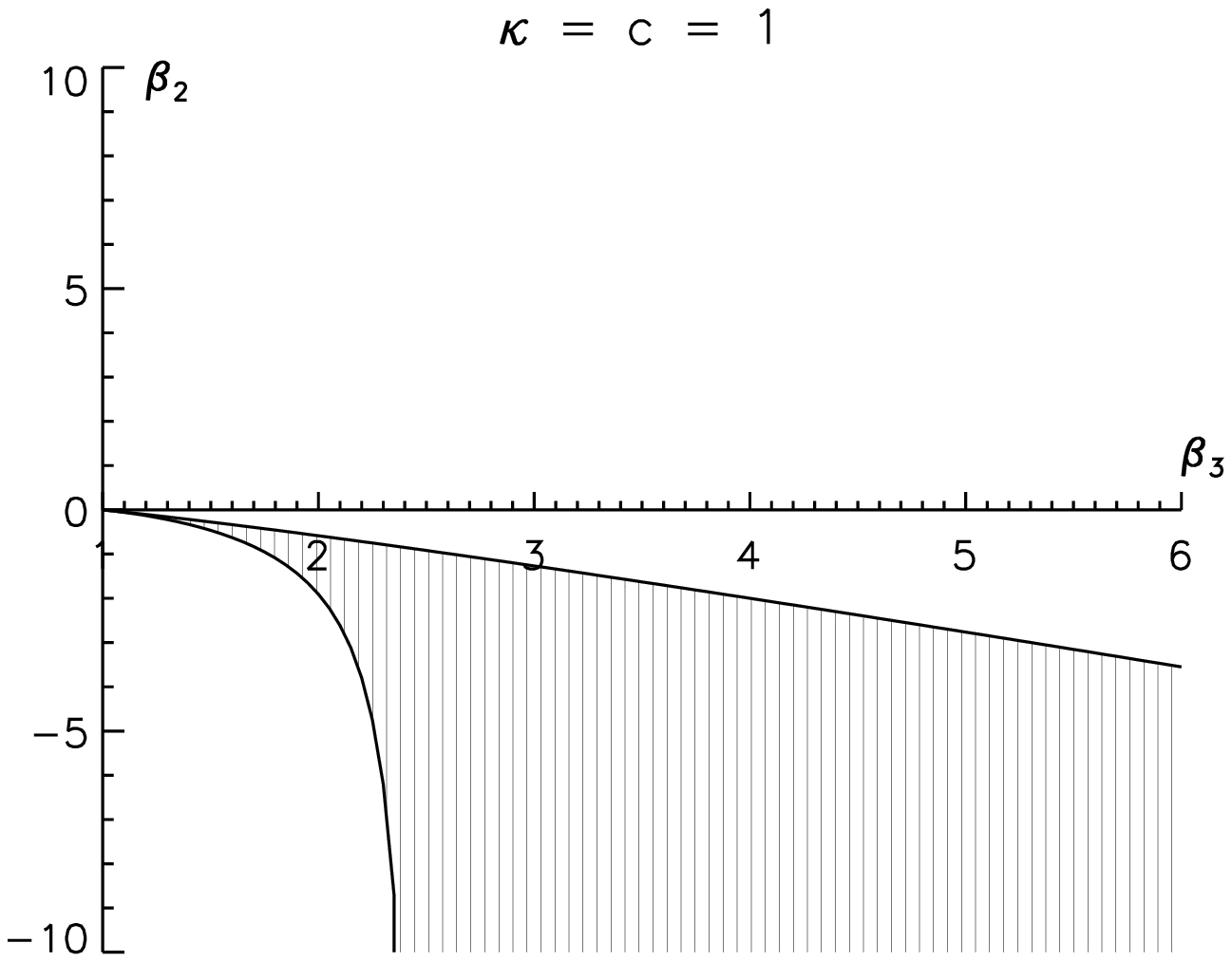}
\par\end{centering}
\caption{\label{fig:ineq}{\em Left panel:}  Allowed region (hatched) for $\alpha$ and $\beta$ using $\kappa=c=1$. {\em Right panel:}  Allowed region (hatched) for $\beta_2$ and $\beta_3$ using a normalization where $\beta_1c+2\beta_2c^2+\beta_3c^3=1$ and $\kappa=c=1$.}
\end{figure}

In Sec.~\ref{app:real}, we show that real valued solutions to Eq.~\ref{eq:poly} that approach zero at infinity (which corresponds to the asymptotically flat solutions) exist if $\alpha>-1/\sqrt{\beta}$. Furthermore, one must also have $\alpha<-d_1/d_2$  when $d_2<0$, where
\begin{eqnarray}
d_1 &\equiv & 1+3\kappa c^{2}-6\sqrt{\beta}(1+\kappa c^{2})+3\beta(1+\kappa c^{2}),\nonumber \\
d_2 &\equiv & -1+6\sqrt{\beta}(1+\kappa c^{2})(1+\beta)-\beta(13+12\kappa c^{2}).
\end{eqnarray}
These constraints are depicted in Fig.~\ref{fig:ineq} and are more restrictive than those presented in Ref.~\cite{Babichev:2013pfa}.
In terms of $\beta_i$, we can write $\alpha>-1/\sqrt{\beta}$ and $\beta >1$ as (using the normalization where $\beta_1c+2\beta_2c^2+\beta_3c^3=1$)
\begin{eqnarray}\label{eq:vainbound}
\beta_3 c^3 &>& 1,\nonumber \\
\beta_2 c^2 &<& \sqrt{\beta_3 c^3}-\beta_3 c^3<0,\nonumber \\
\beta_1 c &=& 1-2\beta_2 c^2-\beta_3 c^3>(\sqrt{\beta_3 c^3}-1)^2>0.
\end{eqnarray}
That is, we need $\beta_2$ to be strictly negative and $\beta_1$ and $\beta_3$ to be strictly positive.

For the phenomenological analysis, we will use the following definitions of the potentials:
\be\label{eq:potdef}
\Phi \equiv Q^2-1\simeq 2\delta Q,\qquad \Psi \equiv 1-N^{-2}\simeq 2\delta N.
\ee
From Eqs.~\ref{eq:deltaQp}, \ref{eq:deltaN} and \ref{eq:potdef}, we have
\begin{eqnarray}
\Psi &=& -\frac{2M\left(r\right)}{r}-\frac{m_g^{2}\kappa c^2 r^{2}}{1+\kappa c^2}\left(\mu-\alpha\mu^2+\frac{\beta\mu^{3}}{3}\right) ,\\
v^2&=& r\Phi^\prime = \frac{P\left(r\right)}{M_g^2}r^{2}+\frac{2M\left(r\right)}{r}-\frac{m_g^{2}\kappa c^2 r^{2}}{1+\kappa c^2}\left(\mu-\frac{\beta\mu^{3}}{3}\right),
\end{eqnarray}
where $v$ is the circular velocity.
From this we see that as long as $\mu$ stays real and finite as $r\to 0$, we will recover GR at small radii, as long as the potentials are small.

Outside the source, we have 
\begin{eqnarray}\label{eq:outside}
\Psi &=& -\frac{2\Mt}{r}\left[1+\frac{\kappa c^2}{1+\kappa c^2}\left(\frac{r}{r_V}\right)^3\left(\mu-\alpha\mu^2+\frac{\beta\mu^{3}}{3}\right)\right] \\ \nonumber
&\xrightarrow[]{r\gg r_V}& -\frac{2\Mt}{r}\left[1-\frac{\kappa c^{2}}{3\left(1+\kappa c^{2}\right)}\right],\\
 r\Phi^\prime & =& \frac{2\Mt}{r}\left[1-\frac{\kappa c^2}{1+\kappa c^2}\left(\frac{r}{r_V}\right)^3\left(\mu-\frac{\beta\mu^{3}}{3}\right)\right]\\
 &\xrightarrow[]{r\gg r_V}&\frac{2\Mt}{r}\left[1+\frac{\kappa c^{2}}{3\left(1+\kappa c^{2}\right)}\right].
\end{eqnarray}
The limit $r\gg r_V$ is derived by noting that we can neglect higher order terms in $\mu$, and solve Eq.~\ref{eq:poly} as 
\begin{equation}
\mu=-\frac{1}{3}\left(\frac{r_V}{r}\right)^3.
\end{equation}
Note that this ``asymptotic'' value is only valid far inside the Compton wavelength of the graviton and represents the maximal deviation we expect from GR.  The deviation 
is monotonically increasing with $\kappa c^2$ and has a maximal value of $1/3$. For $r\gg \lambda_g$, we recover GR again. 
If we are far outside the graviton Compton wavelength, the exponential term will be negligible and GR is recovered. If we are far inside the Vainshtein radius $r_V\sim (\Mt \lambda_g^2)^{1/3}$, and GR is restored again. We thus expect the largest deviations from GR to happen at $ r_V\lesssim r\lesssim \lambda_g$. As an example, assuming $m_g=H_0$, for the Sun, we have 
\be
r_V^\odot\approx 4\cdot 10^{18}\,{\rm m}\approx 140\,{\rm pc}\approx 2.7\cdot 10^7\,{\rm AU}.
\ee
We thus only expect the gravitational field from the Sun to be modified on scales much larger than the distances to its closest star neighbours, where they of course are completely negligible anyway. In the solar system ($r\sim 1\,{\rm AU}$), deviations are of order $(r/r_V)^3\approx 10^{-21}$. This value is way below current observational constraints showing that on AU scales, deviations from the inverse square force law is $\lesssim 10^{-9}$ \cite{2014LRR....17....4W}.
This observational constraint indicate that $\lambda_g\gtrsim 1\,{\rm kpc}$, except for the case of $\kappa c^2\ll 1$ when constraints on $\lambda_g$ will be weaker. 

For parameter values not fulfilling the requirements given, we may still have everywhere real solutions if the radius of the source is bigger than its Vainshtein radius. For example, if the source has a constant density (and zero pressure), $\mu$ will be constant inside the source. As shown in Ref.~\cite{Mortsell:2015exa}, the requirement of having sources larger than their Vainshtein radii corresponds to them having densities smaller than order of the critical density of the Universe, if $m_g\sim m \sim H_0$.  We still might be able to have more compact sources if we let $\kappa$ be very small since $m_g\propto m\kappa^{-1/2}$. 
For a source with mean density $\rho_\star$, for small $\kappa$, we can write Eq.~\ref{eq:poly} at the surface of the source 
 \begin{align}
3\mu+6\left(1-\alpha\right)\mu^{2}+\frac{1}{3}\left[6\alpha^{2}-34\alpha+4\beta+10\right]\mu^{3}+\nonumber \\
\frac{2}{3}\left[6\alpha^{2}-7\alpha+4\beta+1\right]\mu^{4}+
\frac{1}{3}\left[2\alpha^{2}-\beta^{2}+2\beta-4\alpha\beta-2\alpha\right]\mu^{5} \nonumber \\
=-\left(\frac{\rho_\star}{\rho_{\rm cr}}\right)\left(\frac{H_0}{m_g}\right)^2\left(1+\mu\right)^{2}\left(1-\beta\mu^{2}\right),
\end{align}
where the critical background density of the Universe today is given by
\be
\rho_{\rm cr} =  3M_g^2H_0^2\sim 1.88\cdot 10^{-29}\,h^2\,{\rm g\,cm}^{-3}.
\ee
Demanding that solutions exist down to the surface of a neutron star for which 
\be
\rho_{\rm neutr} \sim  4\cdot 10^{14}\,{\rm g\,cm}^{-3},
\ee
we generally need 
$\left(\rho_\star/\rho_{\rm cr}\right)\left(H_0/m_g\right)^2\approx \left(\rho_\star/\rho_{\rm cr}\right)\kappa\left(H_0/m\right)^2$ to be smaller than order one, which means that $\kappa$ should be less than the ratio of the critical density of the universe and the source density, assuming $m\sim H_0$.\footnote{Although fields are not weak at the surface of a neutron star, using the polynomial equation is sufficient to obtain order-of-magnitude estmates.}
For general values of $m$ we get
\begin{equation}
\kappa\lesssim 10^{-44}\left(\frac{m}{H_0}\right)^2.
\end{equation}
Alternatively, we can constrain the Compton wavelength of the graviton
\begin{equation}
\lambda_g\lesssim \sqrt{\frac{\rho_{\rm cr}}{\rho_{\rm neutr}}}r_H\simeq 28\,{\rm km}.
\end{equation}
where  $r_H\equiv H_0^{-1}\approx 1.3\cdot 10^{26}\,{\rm m}$. Note however that this very restrictive limit only needs to be
fulfilled for parameter values not fulfilling the ones illustrated in Fig.~\ref{fig:ineq}.

In Ref.~\cite{2015arXiv150307521A} a limit of $\sqrt{\kappa}\lesssim 10^{-17}$ was derived in order to push scalar instabilities back before BBN. 
For this to work, we need at least two $\beta_i\neq 0$.
For background and perturbation solutions, the main idea is that in the limit of $\kappa\to 0$, the ratio between the scale factors of the two metrics goes to a constant determined by the values of the $\beta_i$ and $c$. This gives a cosmological constant-like contribution to the Friedmann equation and well-behaved perturbation theory. 

\section{Galaxies}
\label{sec:galaxies}

We now turn to the phenomenology of galaxies. In the dark matter paradigm, there is a somewhat unexpected large correlation between the distribution of baryonic and dark matter. One of the main arguments for MOND is that it is able to explain this correlation on galactic scales \cite{Milgrom:1983ca}. However, it fails on larger scales \cite{Pointecouteau:2005mr}. The Vainshtein radius on the other hand naturally adapts to the scale of the object.

\begin{figure}
\begin{centering}
\includegraphics[scale=0.65]{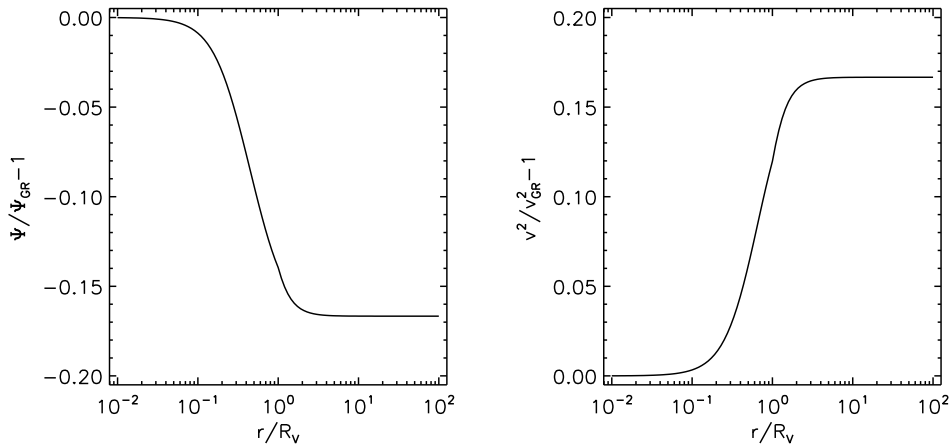}
\par\end{centering}
\caption{\label{fig:Vainshtein} Deviations from the general relativity predictions for the potential $\Psi$ and the circular velocity $v^2$ for the case of $\kappa=c=l=1$ (where $l=r_g/r_V$), $\alpha=1$ and $\beta=4$.}
\end{figure}

We add a galactic source (with negligible pressure) with density profile
\begin{equation}
\rho (r)=\rho_0r^{-q},
\end{equation}
being truncated at $r=r_g\equiv l r_V$, where the parameter $l$ sets the compactness of the galaxy and is of order one, or slightly lower, for $m_g\sim H_0$. We then have
\begin{align}
M(r)=& \rho_0\frac{r^{3-q}}{2(3-q)},\\ \nonumber
\Mt =& \rho_0\frac{r_g^{3-q}}{2(3-q)}.
\end{align}
We can now write (inside the galaxy; outside the galaxy the solution is given by Eqs.~\ref{eq:outside})
\begin{eqnarray}
\Psi &=& -\frac{2M(r)}{r}\left[1+l^{3-q}\frac{\kappa c^2}{1+\kappa c^2}\left(\frac{r}{r_V}\right)^{-q}\left(\mu-\alpha\mu^2+\frac{\beta\mu^{3}}{3}\right)\right] ,\\ \nonumber
v^2&=& r\Phi^\prime = \frac{2M(r)}{r}\left[1-l^{3-q}\frac{\kappa c^2}{1+\kappa c^2}\left(\frac{r}{r_V}\right)^{-q}\left(\mu-\frac{\beta\mu^{3}}{3}\right)\right].
\end{eqnarray}
This typically gives result like in Fig.~\ref{fig:Vainshtein} where $\kappa=c=l=1$, $\alpha=1$ and $\beta=4$.

The observed gravitational lensing and dynamical properties of elliptical galaxies are consistent with general relativity predictions, to an accuracy of $\sim 5\,\%$ \cite{2010ApJ...708..750S,Enander:2013kza}.  
We have three ways to make our model consistent with lensing constraints. The first is to make the Compton wavelength so small that we are well outside it for the lensing and dynamical observations (basically, the velocity dispersion of stars). The lensing radii typically are $\simeq 5$ kpc and the velocity dispersion integrated out to similiar radii. In order not to be in conflict with the observed constraints, we thus need $\lambda_g\lesssim 0.5$ kpc. However, as noted in Sec.~\ref{sec:stars}, such small values are ruled out by Solar system constraints. 
 
The second possibility is that the so called gravitational slip $\gamma$ -- the ratio of the gravitational potentials experienced by massive and massless particles -- is small.
The largest deviations from general relativity predictions are found between the Vainshtein radius and the Compton wavelength where
\begin{equation}
\gamma=\frac{\kappa c^{2}}{3(1+3\kappa c^{2})}
\end{equation}
Using data from the strong gravitational lens sample observed with the Hubble
Space Telescope Advanced Camera for Surveys by the Sloan Lens ACS (SLACS) Survey \cite{Bolton:2008xf}, we 
constrain $\kappa c^2\lesssim 0.1$ at $2\,\sigma$.

The third possibility, valid for parameter values for which we have a functioning Vainshtein mechanism, is to make sure that we are well inside the Vainshtein radius,
\be
r_V=\left(\frac{2M_g}{m_g^2}\right)^{1/3}\simeq 630\left(\frac{M}{10^{11}\,M_\odot}\right)^{1/3}\left(\frac{H_0}{m_g}\right)^{2/3}\,{\rm kpc}.
\ee
For large $\kappa c^2$, we typically need to be a factor of 10 inside the Vainshtein radius not to be in conflict with observational limits, corresponding to $m_g/H_0\lesssim 40$ or $\lambda_g\gtrsim 0.1\,{\rm Gpc}$.

To summarize, strong lensing galaxy systems constrain the graviton Compton wavelength $\lambda_g$ to be either smaller than $\sim 0.5$~kpc or larger than $\sim 0.1$~Gpc, {\em or} the combination $\kappa c^2$ to be smaller than $\sim 0.1$. However, $\lambda_g<0.5\,{\rm kpc}$ is disfavoured by Solar system constraints. We also note that we generally expect the velocity dispersion in galaxies and galaxy clusters to increase as compared to the general relativity prediction, on scales similar to the sizes of the systems if $m_g\sim H_0$ or slightly larger. This will have an effect on the predicted abundance of dark matter in these systems, namely that we need less dark matter than in the general relativity case. However, since we maximally expect the velocity dispersion squared to increase by a factor of $1/3$, the effect is not large enough to completely evade the need for dark matter in galaxies and galaxy clusters.   

\section{Vacuum solutions}
\label{sec:vacuum}

In the previous section, we studied stars, galaxies and their phenomenology. In this section we comment on the relationship between the star solutions and vacuum solutions, such as black holes. Our chief interest here is to understand if the bimetric black holes can be the end-state of the gravitational collapse of massive stars.

\begin{figure}
\begin{centering}
\includegraphics[scale=0.5]{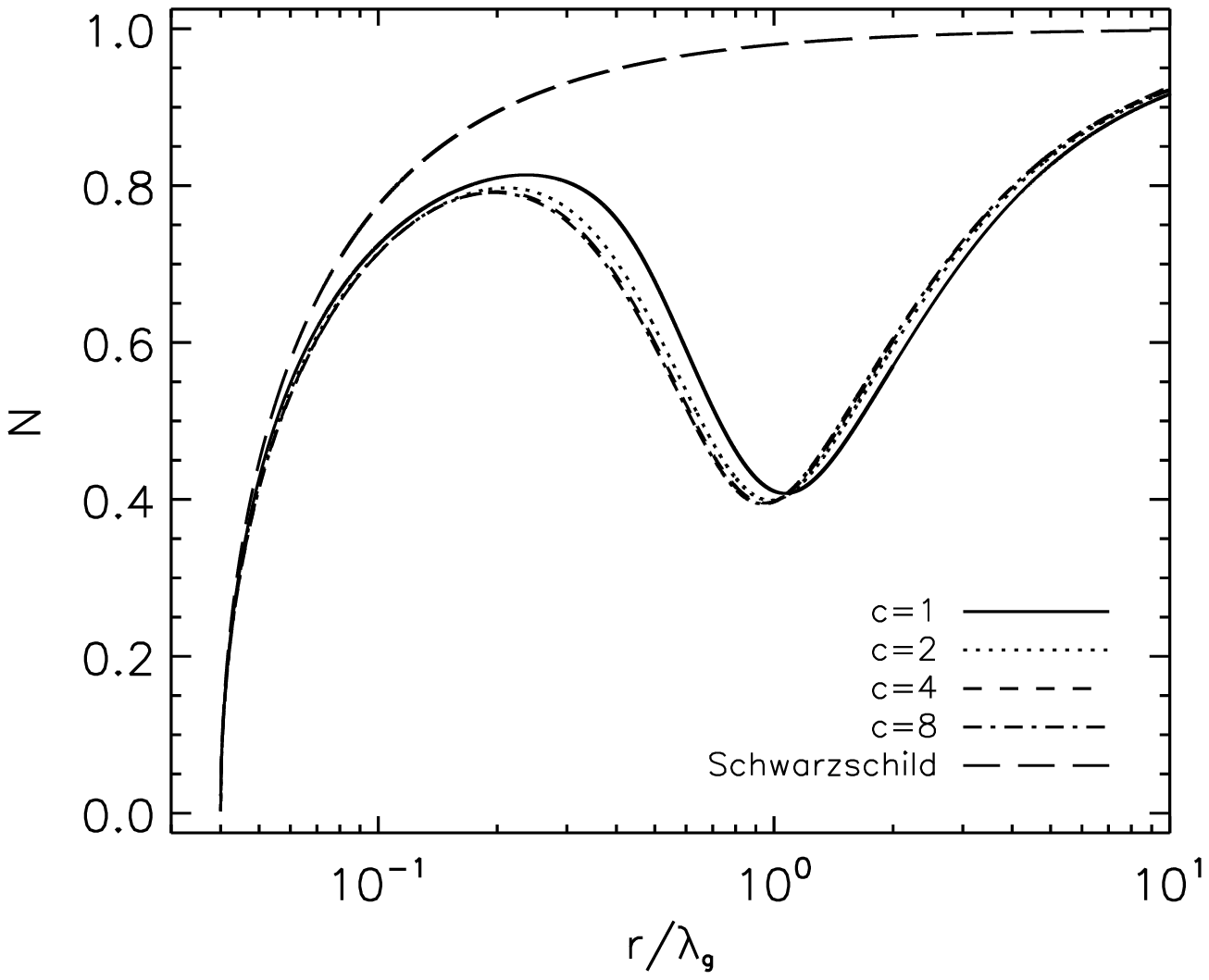}
\includegraphics[scale=0.5]{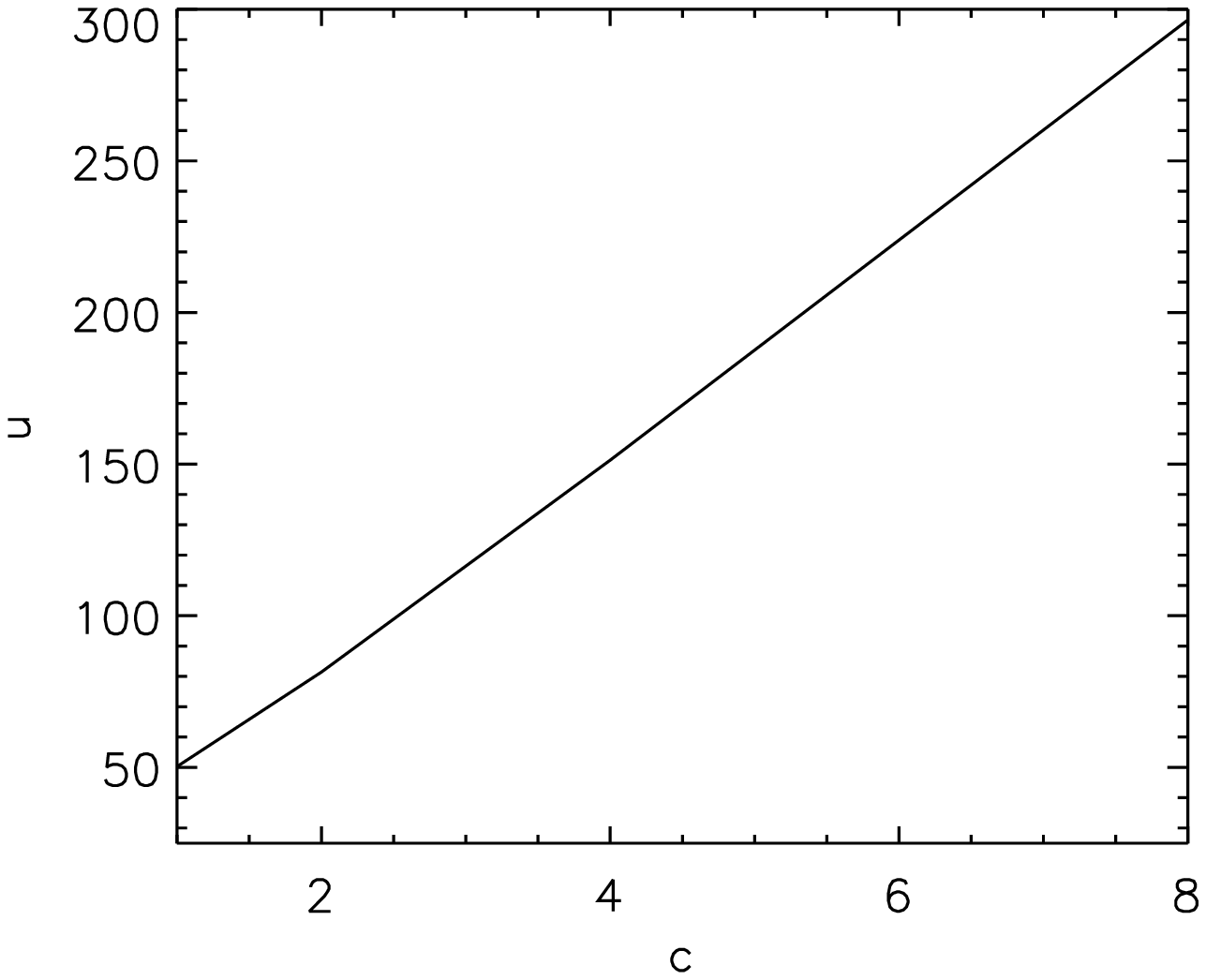}
\par\end{centering}
\caption{\label{fig:bhc}{\em Left panel:} The field $N$ for different choices of $c$, for a black hole with $r_H/\lambda_g=0.04$ and $\beta_2=\beta_3=0$. The fields approach the Schwarzschild solution close to $r_H$. Varying the parameter $c$ shows that several possible solutions exist for a given $r_H/\lambda$. {\em Right panel:} The constant $u$, given by $U/r$ as $r\rightarrow r_H$, as a function of $c$, showing a close to linear relationship.} 
\end{figure}

Vacuum solutions in massive bigravity were studied extensively in Ref.~\cite{Volkov:2012wp}. Following the proof of Ref.~\cite{Deffayet:2011rh}---that for non-singular metrics there has to be a common Killing horizon---we expand the fields $N$, $Y$ and $U$ close to the horizon, situated at $r=r_h$, as
\be
N^2=\sum_{n\geq 1} a_n\left(r-r_h\right)^n,\quad Y^2=\sum_{n\geq1} b_n \left(r-r_h\right)^n,\quad U=ur_h +\sum_{n\geq 1}c_n\left(r-r_h\right)^n.
\ee
From Eqs.~\ref{eq:numsetup}, the coefficients $a_n$, $b_n$ and $c_n$ can all be expressed in terms of $u$ and $a_1$, where $u$ is arbitrary and $a_1$ satisfies a quadratic polynomial with coefficients depending on $u$ and the parameters of the theory (i.e. $c$ and the $\beta_i$ parameters). Since there are three equations of motion, and three free parameters ($u$, $C_1$ and $C_2$) there exists at most a discrete set of solutions for a given value of $c$ and the $\beta_i$ parameters. The structure of these solutions was investigated extensively in Ref.~\cite{Brito:2013xaa}, in the case of $c=1$. This was done through a shooting method, where $u$, $C_1$ and $C_2$ were varied until the solution with asymptotic flatness was found. In this paper we have performed a similar numerical study, but with general $c$. Our results are in agreement with Refs.~\cite{Brito:2013xaa} and \cite{Volkov:2012wp} wherever they overlap.

It was found in Ref.~\cite{Brito:2013xaa}, that for a given value of the $\beta_i$ parameters, the solutions are classified by $r_h/\lambda_g$, i.e the ratio between the horizon and the Compton wavelength of the graviton. An upper bound for $r_h/\lambda_g$ is 0.876, a value related to the Gregory-Laflamme instability (see Ref.~\cite{Volkov:2014ooa} for an interesting discussion of this result). Above that bound, only the bi-Schwarzschild solution exists (i.e. $g_\mn$ is equal to the Schwarzschild solution, and $f_\mn=c^2g_\mn$). The minimum value of $r_h/\lambda_g$ depends on the model under consideration. The conjectured parameter structure presented in Ref.~\cite{Brito:2013xaa} is that when $\beta_3$ is non-zero, solutions cease to exist below a critical value of $r_h/\lambda_g$ (which excludes realistic astrophysical black holes). When $\beta_3=0$, $\beta_2>1$ and $\beta_1<-1$, black hole solutions exist for all values of $r_h/\lambda_g$ below the Gregory-Laflamme bound.

Moving beyond the case of $c=1$, we show in Fig.~\ref{fig:bhc} the metric field $N$ for different values of $c$ and with $r_H/\lambda_g=0.04$. This shows that several possible black hole solutions are possible for fixed $r_H/\lambda_g$, as long as $c$ is varied. We also display the relationship between the constants $u$ and $c$. 

\begin{figure}
\begin{centering}
\includegraphics[scale=0.5]{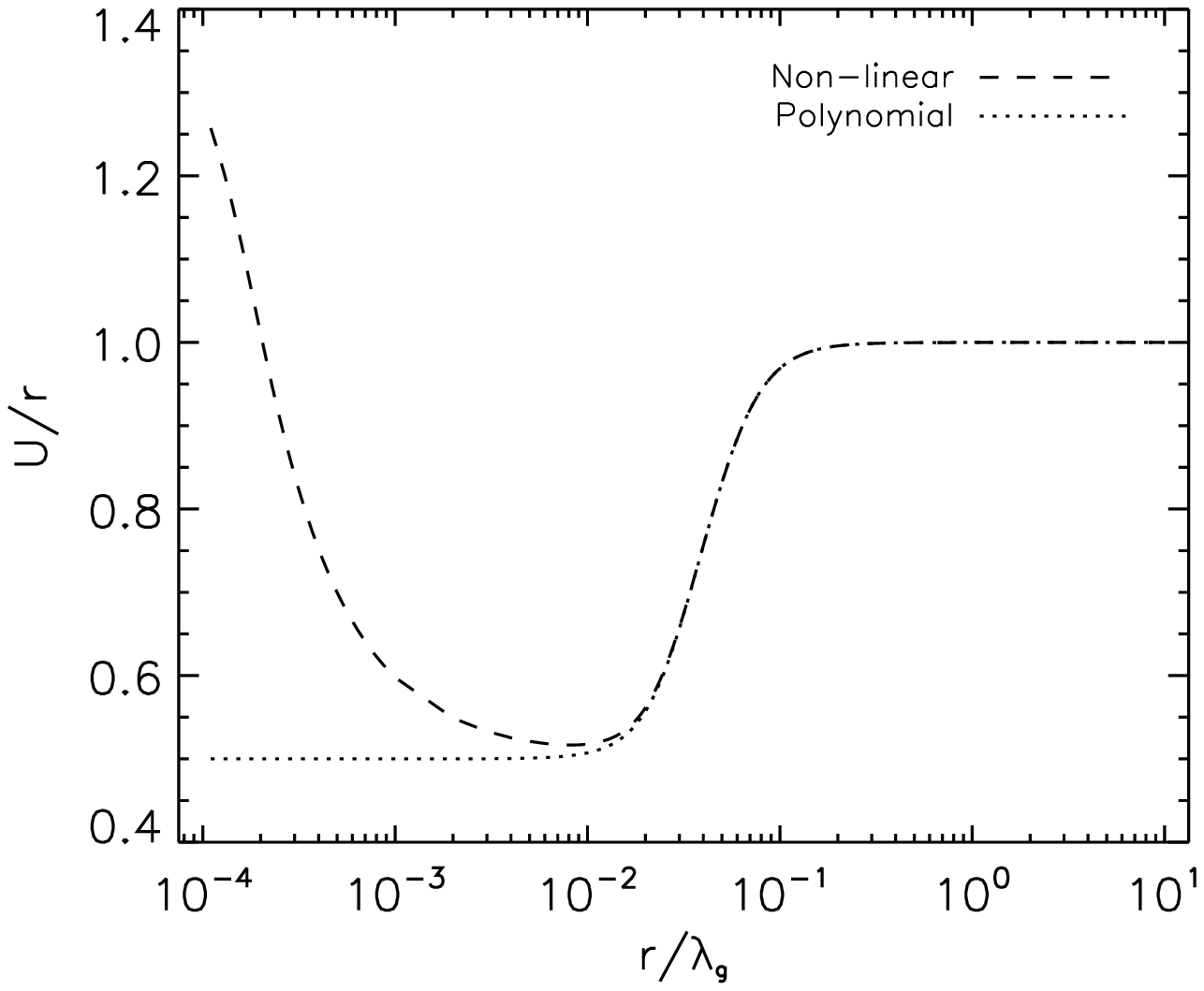}
\includegraphics[scale=0.5]{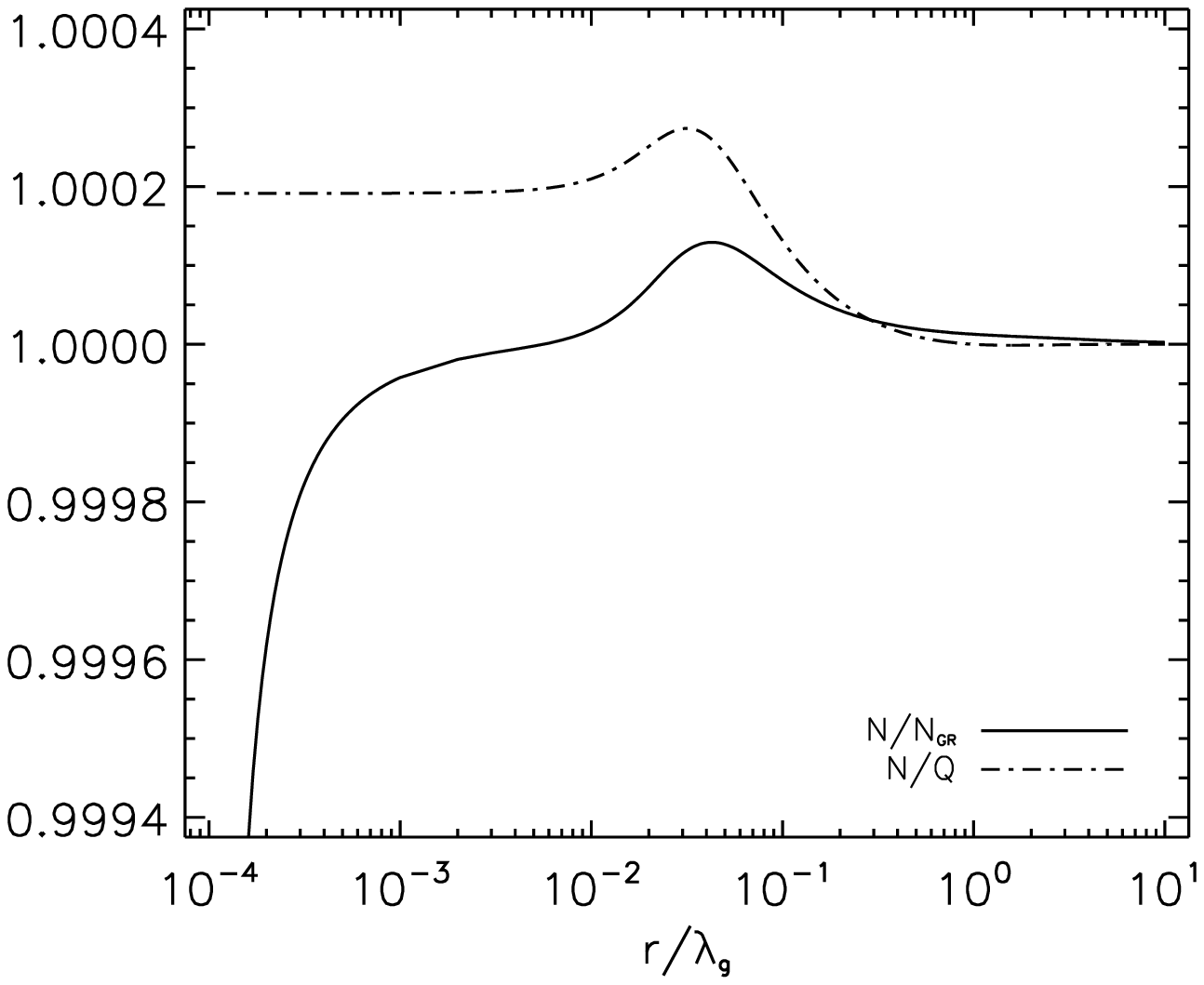}
\par\end{centering}
\caption{\label{fig:vacuum}{\em Left panel:} The function $U/r$ solved using the full equations of motion numerically (dashed) and using the approximate solution, given by Eq.~\ref{eq:poly} (dotted). $U/r$ departs from the constant solution predicted by the approximate solution when the other metric fields become non-linear. {\em Right panel:} The metric function $N$ divided by the GR-solution and $N$ divided by $Q$. For the GR-solution, the Schwarzschild radius is given by $r_S/\lambda_g=10^{-4}$; this ensures that the horizon of the GR and bigravity solutions conicide. In both the left and right panel, $C_1/\lambda_g=5\times 10^{-5}$, $C_2=-2/3\times C_1$ and $\beta_1=7$, $\beta_2=-5$, $\beta_3=4$, $c=\kappa=1$ (these specific values ensure that the solution exists within the Vainshtein radius). $N/N_{GR}$ and $N/Q$ approach unity as $C_1/\lambda_g$ decreases.}
\end{figure}

{\em Stars and black holes.} Concerning the relationship between the star and black hole solutions, we note the following: First of all, $u<c$ for the stars, but $u>c$ for the black holes. Secondly, for star solutions to exist inside the Vainshtein radius, we must have $\beta_3c^3>1$. For the black holes, we must instead have $\beta_3=0$ for solutions to exist for all $r_h/\lambda_g<0.876$, according to the conjecture of Ref.~\cite{Brito:2013xaa}. Finally, the asymptotic structure is different as compared to the black holes and stars. For stars, we have $C_2/C_1 = -2/3$. For the black holes, while a full parameter scan is beyond the scope of this paper, we conjecture that all black holes satisfy 
\be
0\leq\frac{C_2c^2}{C_1}<\frac{2}{3}.
\ee
This conjecture follows from a numerical analysis, where we find that the point $C_2c^2=2C_1/3$ (in the following we put $\kappa=1$) marks a transition for the behaviour of $N$. Above this value, i.e. $C_2c^2>2C_1/3$, $N$ will generically become larger than unity.\footnote{As a sideremark, we note that for $C_2c^2>2C_1/3$, there exist solutions where all the metric fields beside $a$ go like $\sim 1/\sqrt{r}$, inside the Vainshtein radius, for the $\beta_2=\beta_3=0$ model. The implication of these solutions will be investigated in an upcoming work.} Below this value, $N$ will become less than unity. Furthermore, for $C_2>0$, $U/r$ will grow larger than $c$ as one integrates from infinity towards lower $r$, and for $C_2<0$, it will become smaller. The point $C_2=0$ corresponds to the Schwarzschild solution, and as $C_2\rightarrow 0$, $r_h/\lambda_g$ approaches the value 0.876 given by the Gregory-Laflamme instability. For the black holes, we have that $N$ should become less than unity (and eventually approach zero), and $U/r$ should be larger than $c$. Thus, for the black holes, we should have $0<C_2c^2<2C_1/3$. This is also confirmed for the black hole solutions that we have studied. There thus seem to be qualitative difference concerning the overall sign of the massive Yukawa modes when comparing stars and black holes. This stands in contrast to the case of general relativity, where a spherical collapse of a massive star into a black hole does not change the asymptotic spacetime structure.\footnote{It is possible that during the collapse process, information concerning the change of the asymptotic structure could propagate outwards at a finite speed. This information would take an infinite time to change the asymptotic structure, a process which has been observed in general relativity during the collapse of massive scalar fields \cite{Okawa:2014nea}. This could potentially reconcile the different asymptotic structure for stars and black holes. It is an open question whether this is the case in massive bigravity (we thank the referee for pointing this out).}

What happens, then, in vacuum when the asymptotic structure of stars is imposed? Solving the full numerical system, we find that the fields of $g_\mn$ approach the Schwarzschild solution (for parameters that satisfy the bounds given in Eq.~\ref{eq:vainbound}). The function $U/r$ remains constant in a region inside the Vainshtein radius, but starts to grow close to the horizon of $g_\mn$. The fields $a$ and $Y$ remain small and finite. We depict this scenario in Fig.~\ref{fig:vacuum}.

An interesting curvature invariant, introduced in Ref.~\cite{Deffayet:2011rh}, is 
\be
I\equiv f_\mn g^\mn = \frac{a^2}{Q^2}+\frac{U^{\prime 2}N^2}{Y^2}+\frac{2U^2}{r^2}.
\ee
This function remains finite for all non-singular metrics, in particular for the black hole and star solutions. For the vacuum solution shown in Fig.~\ref{fig:vacuum}, it does, however, diverge close to the horizon of the $g_\mn$ metric. This is related to the fact that there is no common horizon for both $g_\mn$ and $f_\mn$ when $C_2/C_1=-2/3$ in vacuum. 

{\em Instabilities.} Let us also discuss the instabilities that are present for the bi-Schwarzschild solutions. It was shown in Refs.~\cite{Babichev:2013una,Brito:2013wya} that there exists unstable modes when the horizon radius of the source is less than the Compton wavelength of the graviton. This instability is, however, rather mild, with a timescale equal to the inverse graviton mass. When the latter is of the same size as the Hubble scale today, this means that the instability will require the entire lifetime of the universe to grow significantly. It does, therefore, not have to be important for astrophysical black holes. Intriguingly, the instability was shown to be absent for the non-diagonal bi-Schwarzschild solutions \cite{Babichev:2014oua}, as well as for the partially massless case \cite{Brito:2013yxa}. Now, as was argued in Refs.~\cite{Babichev:2013una,Brito:2013wya,Volkov:2014ooa}, the instability shows that the bi-Schwarzschild solution can not be considered the end-state of a gravitational collapse. It is unclear whether the other black hole solutions, with massive hair, are stable or not. On the whole, then, there are two reasons why the end-state of gravitational collapse is unclear: the instabilities present for the bi-Schwarzschild case (which could also be present for the other black hole solutions), and the different asymptotic structure of stars and black holes.

To summarize, there is a qualitative difference between the star and black hole solutions. The end state of a collapse of a star is therefore uncertain. It could lead to a novel spherically symmetric solution that as of yet has not been discovered. It might lead to a time-dependent solution that does not settle down into a static final state. It seems unlikely, however, that it will lead to the black hole solution that share a common horizon for $g$ and $f$. We therefore conjecture that black holes in massive bigravity can not be formed from the collapse of stars. This is probably due to the fact that the black hole solutions share a symmetry between $g_\mn$ and $f_\mn$ (i.e. a common horizon), whereas the coupling of matter to only one metric, e.g. $g$, breaks this symmetry. An interesting question is whether this conjecture also holds true when coupling matter to both fields.

\section{Conclusions}
\label{sec:conc}

In this paper, we have investigated the phenomenology of stars and galaxies in massive bigravity. Furthermore, we have discussed the relationship between black holes in massive bigravity and stars.

For the stars, we have been interested in the existence of solutions where the radius of the star is much smaller then the Compton wavelength of the graviton. The latter is usually assumed to be of the order of the Hubble scale of the universe today, when massive bigravity is used for cosmological applications. The parameter constraint that we found, which generalizes earlier work in Ref.~\cite{Babichev:2013pfa}, states that $\beta_2$ needs to be strictly negative and $\beta_1$ and $\beta_3$ needs to be strictly positive. If these conditions are not met, we have shown that the ratio  between the Planck masses of the two metrics needs to be less than $10^{-22}$, when the length scale of the theory is of the order the Hubble scale today. 

Moving on to galaxies, we show that the graviton Compton wavelength $\lambda_g$ either has to be so small (less than $\sim$0.5 kpc) so that the massive Yukawa mode does not produce sizable deviations between the lensing and dynamical observations. This is, however, in conflict with Solar system measurements. Another possibility is that $\lambda_g$ is so large that the galaxies fall within the Vainsthein radius. This requires $\lambda_g \gtrsim 0.1$~Gpc. Yet another possibility is that $\kappa c^2\lesssim 0.1$, which makes the deviation in the gravitational slip undetectable.

Finally, an open and interesting question, that deserves further studies, is the end-state of gravitational collapse. In general relativity, the asymptotic structure is unchanged as a star undergoes spherical collapse to a black hole (a fact related to Birkhoff's theorem). In massive bigravity, we find that the asymptotic structure of stars and black holes is qualitatively different. This is related to the sign of the massive Yukawa mode. This makes it unlikely that the black hole solutions are end-states of gravitational collapse. This could potentially be related to the fact that for the black holes $g_\mn$ and $f_\mn$ have a common Killing horizon. This symmetry is, however, broken by stars, since only one of the metrics couple to matter. It would therefore be interesting to investigate the star solutions when coupling both metrics to matter. 

\appendix

\section{Real solutions}\label{app:real}

In this Appendix we derive constraints on parameter values needed to have static, spherically symmetric solutions that are asymptotically flat and valid att all radii. We will make numerous references to the left hand side (LHS) and right hand side (RHS) of the polynomial equation~(\ref{eq:poly}). Assuming that we are outside the source, $M(r)=\Mt$, the pressure is zero and we start by noting that the the RHS is zero at $\mu=-1$ and $\mu=\pm 1/\sqrt{\beta}$ (for $\beta >0$; for $\beta\le 0$, the only root is at $\mu =-1$) and is being divided by $r^3$. As $r\to\infty$, the RHS becomes flat and as $r\to 0$, ${\rm RHS}\to\pm\infty$, except at the points where it is zero. 
Defining (for the pressureless case)
\begin{equation}
h = -\frac{{\rm RHS}}{1+\kappa c^{2}}\left(\frac{r}{r_V}\right)^3=\left(1+\mu\right)^{2}\left(1-\beta\mu^{2}\right)
\end{equation}
we have 
\begin{equation}
\frac{dh}{d\mu} =2\left(1+\mu\right)\left(1-2\beta\mu-3\beta\mu^{2}\right),
\end{equation}
which is zero at $\mu =-1$. Furthermore, 
\begin{equation}
\frac{d^2h}{d\mu^2} =2\left(1-2\beta-10\beta\mu-9\beta\mu^{2}\right),
\end{equation}
which at $\mu =-1$ is $2(1-\beta)$. This is negative for $\beta>1$ and vice versa.

The LHS on the other hand has a shape that is fixed by the values of $\beta_i,\mu,\kappa$ and $c$. It is always zero at $\mu=0$ and 
\begin{equation}
\frac{d{\rm LHS}}{d\mu}\biggr\rvert_{\mu=0}=3\left(1+\kappa c^{2}\right)>0.
\end{equation}
Since we want our solutions to be asymptotically flat we need $\mu\to 0$ as $r\to\infty$. The limiting value for $\mu$ is either $\mu =-1$ if $\beta\le 1$ or $\mu=-1/\sqrt{\beta}$ if $\beta>1$. We first show that $\beta\le 1$ is not an option. For $\mu =-1$, the LHS becomes
\begin{equation}
{\rm LHS}=\frac{1}{3}(1+2\alpha+\beta)^2.
\end{equation}
This has to be smaller than or equal to zero in order for solutions not to become imaginary as $r\to 0$, since the RHS gets arbitrarily negative close to $\mu=-1$. This means that we need to set $\beta = -1 -2\alpha$. Close to $\mu=-1$, we can expand the RHS and LHS sides as
\begin{eqnarray}
{\rm LHS} &\sim & -\frac{2}{3}(1+\alpha)\left[3+2\kappa c^{2}+\alpha(3+\kappa c^{2})\right](\mu +1)^2 ,\nonumber \\
{\rm RHS} &\sim & -2(1+\kappa c^2)(1+\alpha)\left(\frac{r_V}{r}\right)^3(\mu +1)^2 ,
\end{eqnarray}
showing that we will not have real solutions as $r\to 0$ since the RHS always will be less than the LHS for some finite $r$.

\begin{figure}
\begin{centering}
\includegraphics[scale=0.5]{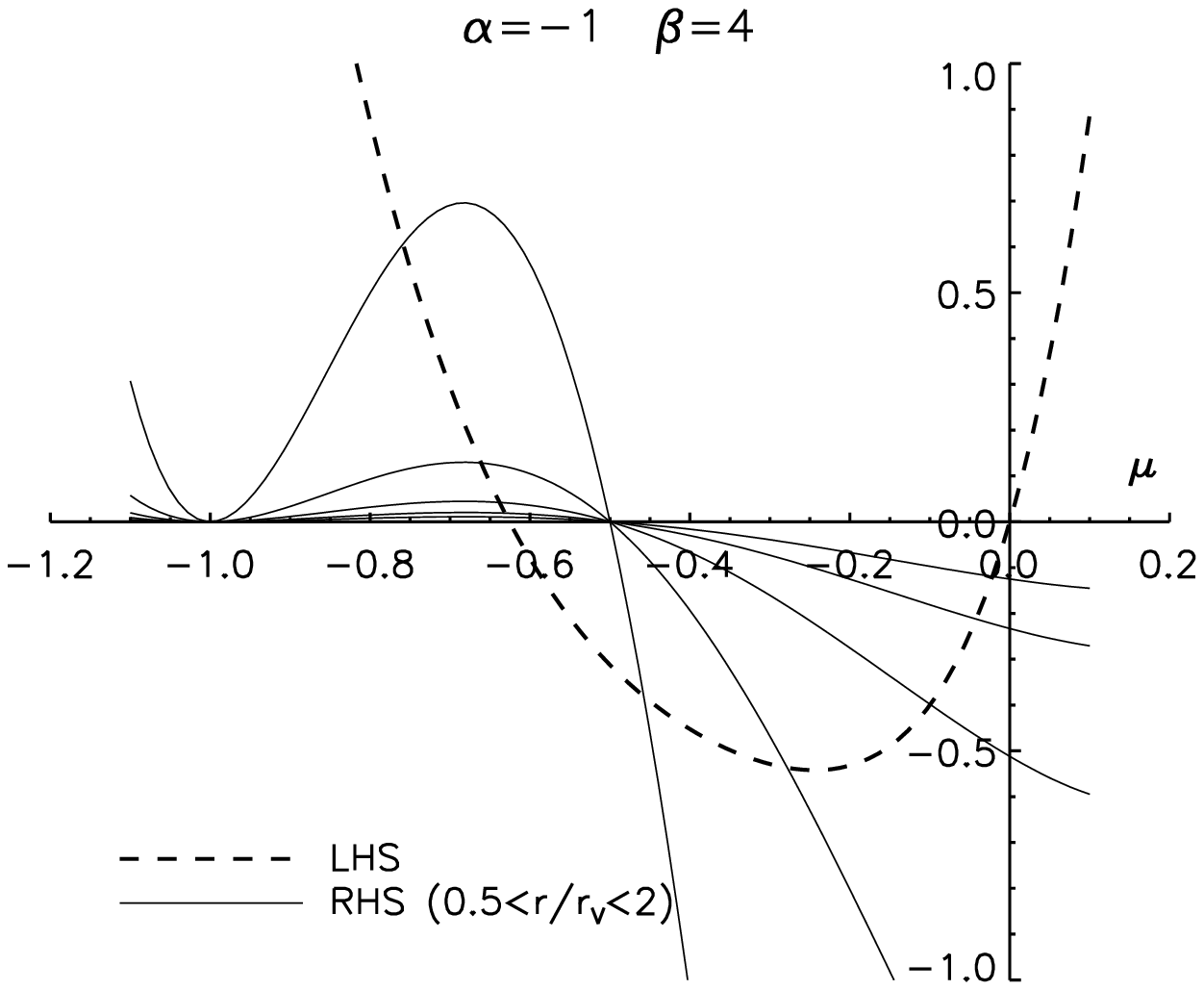}
\includegraphics[scale=0.5]{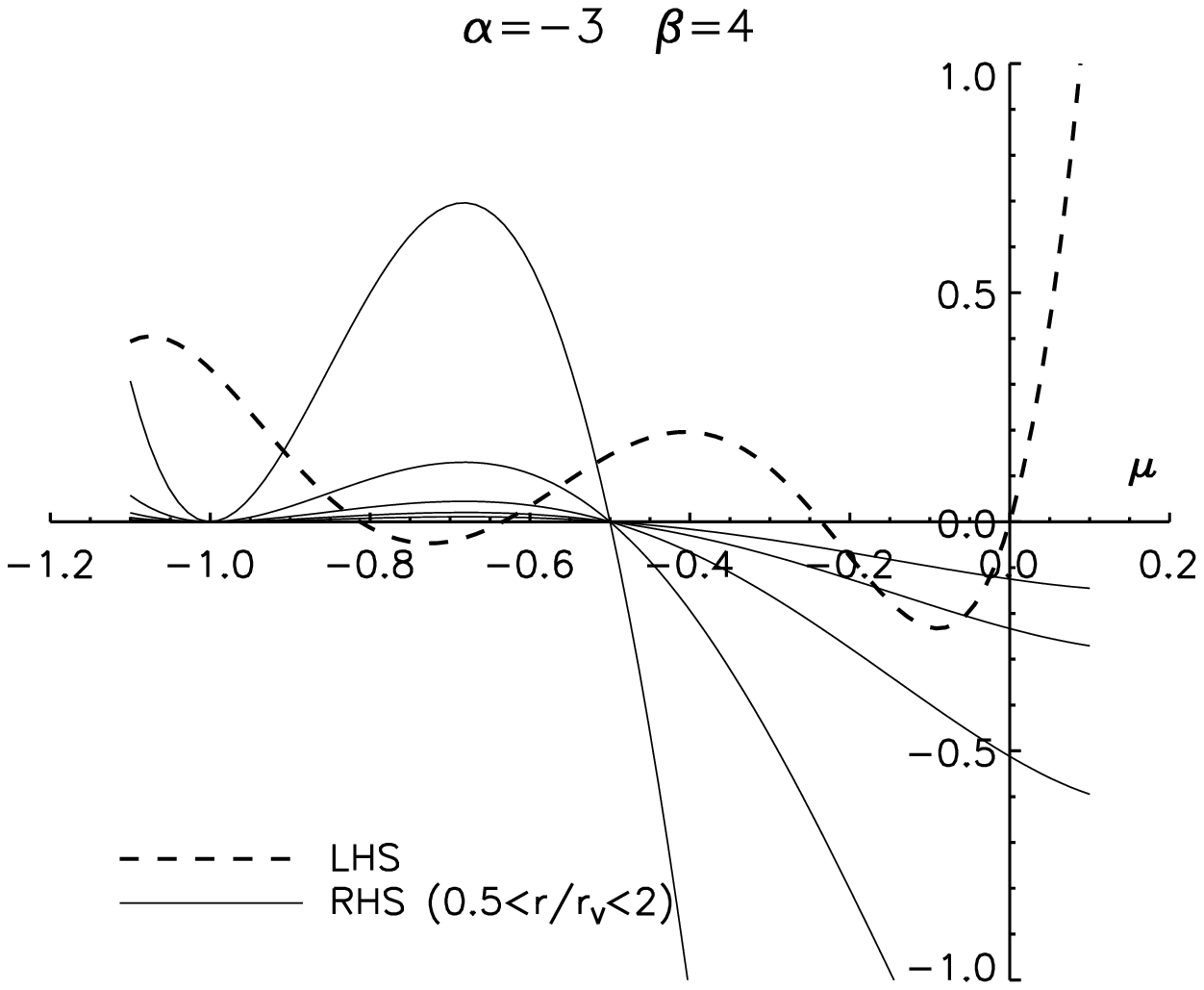}
\par\end{centering}
\caption{\label{fig:alpha-3beta4} Graphic solution for $\kappa=c=1$, $\beta=4$ and $\alpha=-1$ (left panel) and $\alpha=-3$ (right panel).  If the LHS has a maximum or inflection point over the interval $\mu=[-1/\sqrt{\beta},0]$, solutions for which $\mu\to 0$ as $r\to\infty$ will become complex at some finite value of $r$ as illustrated in the right panel. This rules out regions where very low values of $\alpha$.}
\end{figure}

For $\beta>1$, the question is whether we have a real solution for which $\mu=[-1/\sqrt{\beta},0]$ for $r=[0,\infty]$.
For $\mu=-1/\sqrt{\beta}$, we can write the LHS as
\begin{eqnarray}
{\rm LHS}&=&-\frac{2d_2}{3 \beta^{5/2}}(\alpha+\sqrt{\beta})\left(\alpha+\frac{d_1}{d_2}\right),\nonumber \\
d_1&\equiv & 1+3\kappa c^{2}-6\sqrt{\beta}(1+\kappa c^{2})+3\beta(1+\kappa c^{2}),\nonumber \\
d_2&\equiv & -1+6\sqrt{\beta}(1+\kappa c^{2})(1+\beta)-\beta(13+12\kappa c^{2}).
\end{eqnarray}
To have ${\rm LHS}<0$, we need 
\begin{eqnarray}
{\rm sign}(\alpha+\sqrt{\beta})&=&{\rm sign}\left(\alpha+\frac{d_1}{d_2}\right),\quad d_2>0\nonumber \\
{\rm sign}(\alpha+\sqrt{\beta})&\neq&{\rm sign}\left(\alpha+\frac{d_1}{d_2}\right),\quad d_2<0.
\end{eqnarray}
We also need the LHS not to have a maximum or inflection point over the interval $\mu=[-1/\sqrt{\beta},0]$. This rules out regions where $\alpha<-d_1/d_2$ when $d_2>0$ as illustrated in Fig.~\ref{fig:alpha-3beta4}. 
 We are thus left with $\alpha>-1/\sqrt{\beta}$, with the additional constraint $\alpha<-d_1/d_2$ for regions where $d_2<0$, see Fig.~\ref{fig:ineq}. 

\acknowledgments

We would like to thank Marco Crisostomi, Fawad Hassan and Richard Brito for useful discussions.Furthermore, we thank an anonymous referee for several helpful comments that improved the manuscript. E.M. acknowledges support for this study by the Swedish Research Council.

\bibliographystyle{JHEP}
\bibliography{bibliography}{}

\end{document}